\documentclass[a4paper]{article}

\usepackage{url}
\usepackage{graphicx,array}
\usepackage{caption}
\usepackage{subcaption}
\usepackage{xspace}
\usepackage{listings}
\usepackage{todonotes}

\usepackage{mdframed}
\usepackage{multicol}

\newcommand{\umons}{\textsuperscript{$\star$}}
\newcommand{\cetic}{\textsuperscript{$\dagger$}}

\newcommand{\jdbc}{\textsf{jdbc}\xspace}
\newcommand{\jpa}{\textsf{jpa}\xspace}
\newcommand{\hbm}{\textsf{hbm}\xspace}

\newcommand{\sect}[1]{Section~\ref{#1}}

\definecolor{darkgreen}{rgb}{0,0.5,0}
\definecolor{darkblue}{rgb}{0,0,0.5}
\definecolor{darkred}{rgb}{0.5,0,0}

\newcommand{\etal}{et al.\xspace}
\newcommand{\ie}{i.e.,\xspace}
\newcommand{\eg}{e.g.,\xspace}

\newcommand{\related}[2]{\mathsf{related}_{P}(#1,#2)}

\newcommand{\profiles}[1]{\mathsf{mp_M}(#1)}

\newcommand{\resid}[0]{\emph{residual}\xspace}
\newcommand{\suppr}[0]{\emph{removed}\xspace}
\newcommand{\compl}[0]{\emph{complemented}\xspace}
\newcommand{\repl}[0]{\emph{replaced}\xspace}

\title{On the Interaction of Relational Database Access Technologies in Open Source Java Projects}

\author{Alexandre Decan\umons, Mathieu Goeminne\umons\cetic~ and Tom Mens\umons\\
~\\
\umons Software Engineering Lab, University of Mons, Belgium\\
Email: \{ first . last \} @ umons.ac.be\\
\cetic Center of Excellence in Information and Communication Technologies, Belgium\\
Email: mathieu.goeminne@cetic.be
}

\begin{document}

\maketitle

\begin{abstract}
This article presents an empirical study of how the use of relational database access technologies in open source Java projects evolves over time. Our observations may be useful to project managers to make more informed decisions on which technologies to introduce into an existing project and when.
We selected 2,457 Java projects on GitHub using the low-level JDBC technology and higher-level object relational mappings such as Hibernate XML configuration files and JPA annotations.
At a coarse-grained level, we analysed the probability of introducing such technologies over time, as well as the likelihood that multiple technologies co-occur within the same project. At a fine-grained level, we analysed to which extent these different technologies are used within the same set of project files.
We also explored how the introduction of a new database technology in a Java project impacts the use of existing ones.
We observed that, contrary to what could have been expected, object-relational mapping technologies do not tend to replace existing ones but rather complement them.
\end{abstract}


\section{Introduction\label{sec:intro}}

As software systems become more and more complex, the effort required for creating new systems and maintaining  existing ones increases over time.
This effort can be reduced by embedding code in reusable libraries that offer services for supporting a particular aspect of the developed system.
For example, for software systems that strongly interact with a \emph{relational database}, numerous technologies (libraries, APIs and frameworks) exist for connecting the program code to the database. 
Understanding how database technologies tend to replace or complement existing ones in software projects can help project managers in choosing the most appropriate technology, and the most appropriate moment of introducing this technology.

The program code can be connected to the database in various ways. In the simplest case, the code will contain embedded database queries (e.g., SQL statements) that will be interpreted by the database management system.
In more complex cases, especially for object-oriented programs, \emph{object-relational mappings (ORM)} will be provided to translate program concepts (e.g., classes, methods and attributes) into  database concepts (e.g., tables, columns and values), so that database elements can be created, read, updated or deleted (CRUD) directly by manipulating object-oriented views.
Despite the fact that ORMs abstract away from technical connection details in order to facilitate software development, some evolution-related problems remain.

The high level of dynamic of current database access technologies makes it hard for a programmer to figure out which SQL queries will be executed at a given location of the program source code, or which source code methods actually access a given database table or column. Conversely, the high level of abstraction provided by the ORMs makes it hard to determine the impact on the program code of changes in the database schema.
In addition, co-evolving the database and the program requires to master multiple languages and technologies.

\medskip
This paper examines how popular technologies are used in open source Java projects for connecting the source code to a relational database. To do so, we focus on three research questions:

$RQ_1$ -- \emph{When and in which order are database technologies introduced in a project?}
We observe that they tend to be introduced very early in the project's lifetime. 
This is expected, since those technologies are typically central components of the projects in which they occur.
We also observe that multiple database access technologies are used in many projects, and that they tend to be used simultaneously. 
Finally, we study which technologies tend to be complemented by other technologies.

$RQ_2$ -- \emph{How does the introduction of a new technology in a project affect the already included ones?}
With this question we wish to understand whether technologies tend to replace existing ones, or rather complement them. In the former case, the introduction of a new technology would decrease the use of the already included technology. In the latter case, the new technology may serve as a catalyst, leading to an increased of the already included technology.

$RQ_3$ -- \emph{To which extent does the introduction of a new technology impact the way in which a project accesses the database?}
This question focuses on the evolution of project files that use a particular technology, after introducing a new database technology in the project: are these files modified in order to benefit from the newly introduced technology? 
For certain pairs of technologies, we found this to be the case. For most pairs of technologies however, existing database-related files do not substantially adopt the latest introduced technology.

\medskip

The remainder of this paper is structured as follows. \sect{sec:sota} presents attempts to methodically analyse and compare similar technologies that can be found in the scientific literature and puts our research in perspective. \sect{sec:methodo} presents the approach we followed for collecting the data required for our empirical study as well as the methodology for analysing it. The next three sections address our research questions. 
\sect{sec:threats} discusses the threats to validity of our study.
\sect{sec:future-work} discusses possible extensions of the presented study, and \sect{sec:conclusion} concludes.


\section{State of the Art\label{sec:sota}}

While the literature on database schema evolution is very large~\cite{Rahm2006}, few authors have proposed approaches to systematically observe how developers cope with database evolution in practice. Sjoberg~\cite{Sjoberg1993} presented a study where the database schema evolution of a large-scale medical application is measured and interpreted. Vassiliadis \etal~\cite{Vassiliadis2015} studied the evolution of individual database tables over time in eight different software systems. 

Several researchers have tried to identify, extract and analyse database \emph{usage} in application programs. The purpose of the proposed approaches ranges from error checking~\cite{christensen2003precise, Gould2004, Sonoda2011}, over SQL fault localisation~\cite{Clark2011}, to fault diagnosis~\cite{Javid2012}. More recently, Linares-Vasquez \etal~\cite{LinaresVasquez2015} studied how developers document database usage in source code. Their results show that a large proportion of database-accessing methods is completely undocumented.

Several empirical studies have analysed the evolution of library and technology usage. Bauer and Heinemann~\cite{DBLP:conf/csmr/BauerH12} were able to identify distinct evolution scenarios for API dependencies in software projects. The gained knowledge may be useful for evaluating opportunities in API migration and evolution. 
Teyton \etal~\cite{DBLP:conf/wcre/TeytonFB12} identified sets of similar libraries in a large corpus of software projects. The obtained results can be used for suggesting alternative libraries to project managers who want to migrate from a library to another one. In~\cite{DBLP:journals/corr/TeytonFPB13} they investigate how and why library migrations occur. They found that library migrations are relatively rare, and projects that have witnessed more than one migration are exceptional. They also observed that migration is generally an atomic change performed by a single developer in a single commit.

\section{Methodology and Data Extraction\label{sec:methodo}}

The empirical study in this paper focuses on open source Java systems. Java is among the most popular programming languages today, and a large number of technologies and frameworks are available to facilitate relational database access from within Java code. The choice for open source systems is motivated by the accessibility of the entire history of the source code in freely accessible version control repositories. 

\subsection{Considered Database Access Technologies}

In previous work~\cite{GoeminnetEtAl2015-ICSME, GoeminnetEtAl2014}, we considered 26 Java relational database technologies that offer a direct means of accessing a relational database and whose presence in a project is identifiable through static analysis. By analysing the import statements in Java files as well as the presence of specific configuration files, we determined the presence of each of these technologies. We performed a survival analysis of the technologies used in order to determine their relative importance over time in the considered projects.

This paper provides a more in-depth study, by looking at the interaction between object-oriented source code and relational databases at a more fine-grained level. We have selected three popular technologies that are representative of a particular way to connect the source code to a database (embedded SQL, external mapping files, and Java annotations):


\subsubsection*{JDBC}

\jdbc\footnote{\url{oracle.com/technetwork/java/javase/jdbc/}} is a low-level technology for connecting Java programs to a database by sending SQL queries directly from within the source code. 
While version 1.1 was released in 1997, there have been regular version upgrades to cope with the evolution of the Java language.
This technology is still intensively used in numerous projects~\cite{GoeminnetEtAl2015-ICSME}, despite the inherently close coupling that is required between the source code and the database schema.

In our study we consider this technology as being associated to a Java source code file if entities belonging to \texttt{java.sql} are imported in this file.

\subsubsection*{Hibernate}

\emph{ORM technologies} rely on a mapping description for associating (object-oriented) source code elements to database elements. They aim to reduce the so-called \emph{object-relational impedance mismatch}~\cite{Ireland2009}. 
The mapping description can take the form of configuration files, placed aside source code files, to express the relations between the considered entities. 
Hibernate is a popular open source Java framework adopting this solution. It was first released in 2001, and provides an abstraction layer on top of \jdbc.
Hibernate has been criticised by many of not being a 100\% transparent data persistence solution. 

In our study we analyse Hibernate\footnote{\url{hibernate.org/}} XML configuration files (denoted by \textbf{\hbm} hereafter), and consider that a Java file relies on Hibernate technology if at least one Hibernate configuration file mentions the Java file as a code entity resource.

\subsubsection*{JPA}

\emph{Annotation-based mapping descriptions} offer an increasingly popular means to express the relations required by ORM engines. With such mappings, Java annotations are used to mark program elements as counterparts of database entities. The \emph{Java Persistence API}\footnote{\url{oracle.com/technetwork/java/javaee/tech/persistence-jsp-140049.html}} (denoted by \textbf{\jpa} hereafter) is the \emph{de facto} Java standard for annotation-based mappings.
\jpa was first released in 2006, and relies on the Java annotation mechanism that was first introduced in Java 5.
We consider this technology as representative for this kind of mapping description.

In our study we consider that a Java file relates to \jpa if the \texttt{Entity}, \texttt{Embeddable}, or \texttt{MappedSuperclass} annotations from package \texttt{javax.persistence} can be found in this file. 

\subsection*{Discussion}

As witnessed by many discussions on Stack Overflow\footnote{see for example \url{stackoverflow.com/questions/}{\color{red}Q}\\ with {\color{red}Q} = 1607819, 2397016, 2560500 or 530215.}, there is no consensus on which of these three technologies is the most appropriate for any given project, as it may depend on many project-related characteristics, technological choices or even personal preferences. 

One should also note that the use of these technologies is not exclusive. A project may use all of these technologies simultaneously. These technologies may even be used together within the same Java source code files.

\subsection{Selected Projects}

In order to obtain a representative project sample, we based our empirical analyses on Java projects belonging the GitHub project corpus proposed by Allamanis and Sutton~\cite{githubCorpus2013}. Among these projects, 13,307 still had an available Git repository on 24 March 2015.

In order to carry out our empirical study, we selected 2,457 projects from this project corpus for which at least one of the commits contained a reference to either \jdbc, \jpa or \hbm. For each selected project, we extracted the existing relations between source code and database entities from the first commit of each week, and we obtained an historical view of all the files that can be related to a particular technology or to a particular framework.

\begin{table}[!htbp]
\footnotesize
   \begin{tabular}{|l|r|r|r|r|}
        \hline
         & mean & stdev & median & max.\\ 
        \hline\hline
        duration (in weeks) & 76 & 121 & 23 & 812\\
        \hline
        \# commits & 1317 & 6013 & 126 & 174,618\\
        \hline
        \# contributors & 12 & 31 & 4 & 1091\\
        \hline
        \# files in HEAD & 1058 & 3549 & 213 & 103,493\\
        \hline
        \# Java files in HEAD & 512 & 1793 & 88 & 46,661\\
        \hline
    \end{tabular}
\caption{Characteristics of the selected projects. HEAD refers to the latest extracted version.\label{tab:characteristics}} 
\end{table}

Table~\ref{tab:characteristics} shows some of the characteristics of the selected projects. The distribution of metrics values is highly skewed, suggesting evidence of a Pareto principle~\cite{Goeminne2011-SQM}. The duration is expressed in weeks between the first and the last commit.


\begin{figure}[!tbp]
\centering 
\includegraphics[width=0.7\columnwidth]{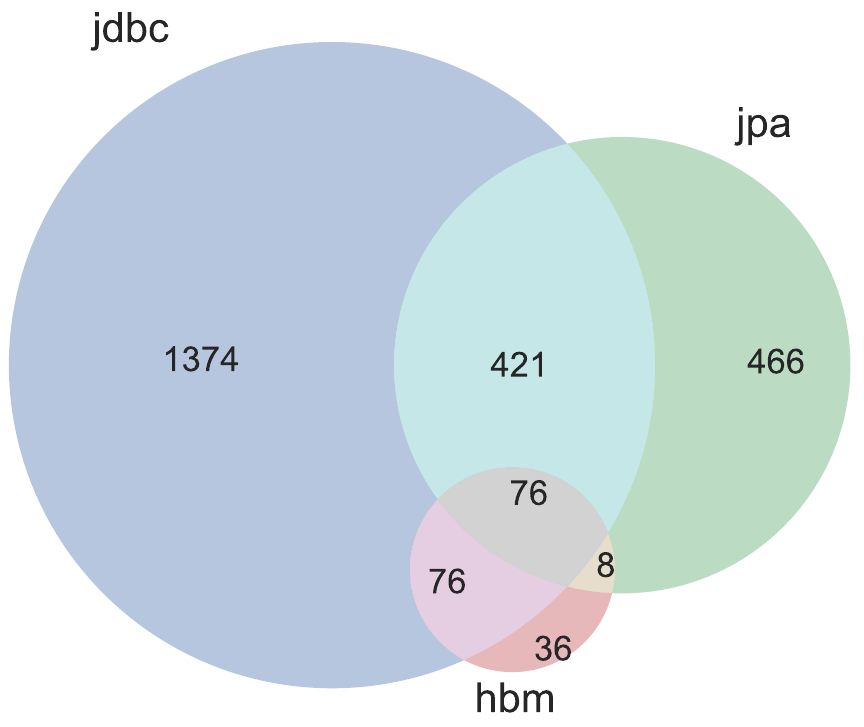}
\caption{Number of projects per considered technology.\label{fig:nb_by_techs}}
\end{figure}

Figure~\ref{fig:nb_by_techs} reports the number of projects per considered technology, taking the entire lifetime of each project into account.
We observe that the project sample is relatively unbalanced with respect to the presence of each technology, but each pair of technologies is still represented in a quite a number of projects.



\section{$RQ_1$ When and in which order are database technologies introduced in a project?}

Introducing a new technology in a software project comes with a certain cost. A common policy is therefore to introduce such a technology only if the expected benefits outweigh the expected cost. 
%

For each project, we analysed at what moment in the projects' lifetime each considered technology got introduced. 
The answer appears to depend on the duration of the considered projects.
To minimise the effect of project duration, we normalised the lifetime of each project into a range between 0 (the start of the project) and 1 (the last considered commit).

\begin{figure}[!htbp]
    \centering
    \includegraphics[width=0.9\columnwidth]{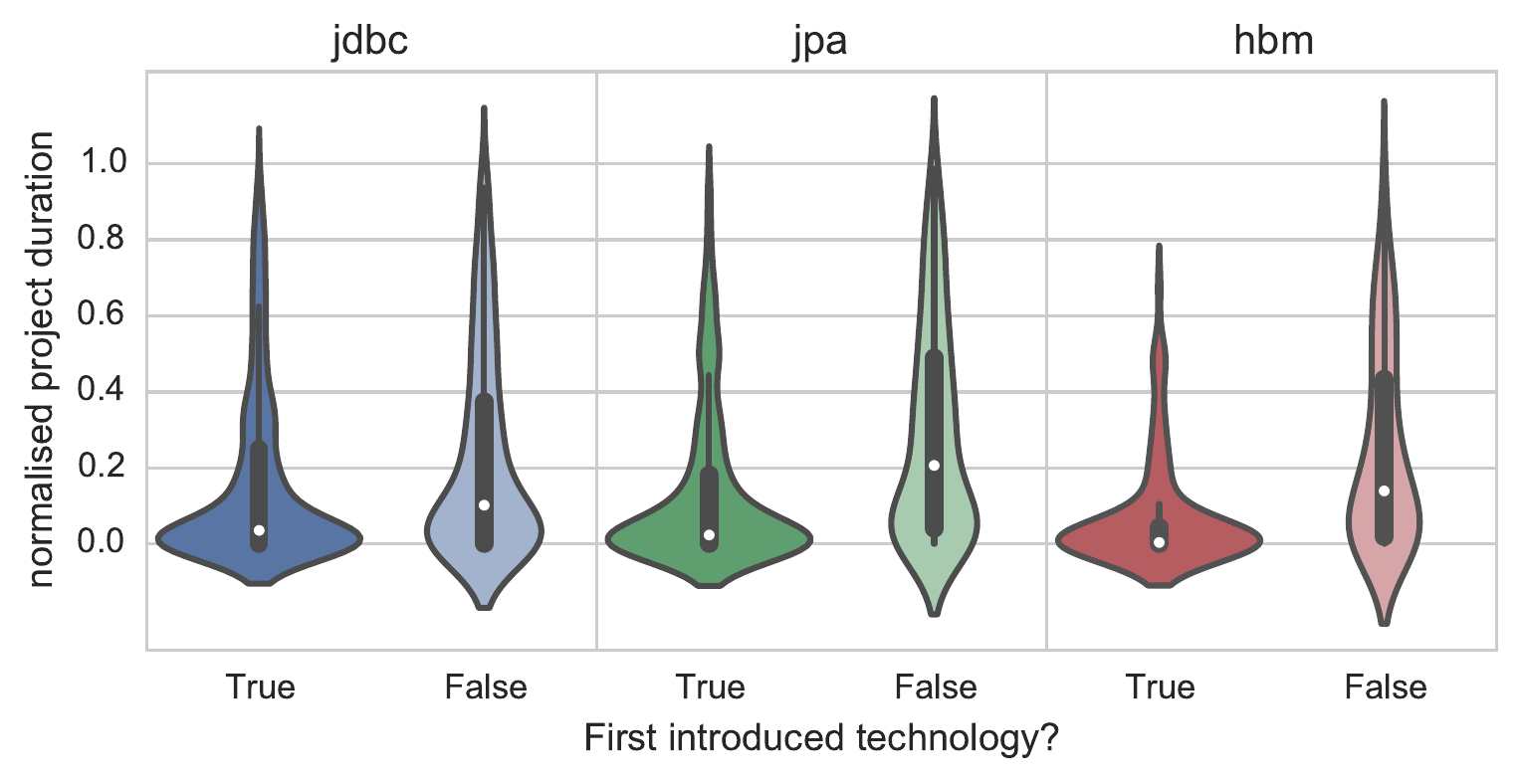}
    \caption{Violin plot (using a kernel density estimate) of the distribution of the introduction time of a technology in the Java project corpus.\label{fig:starttimes}}
\end{figure}

Figure~\ref{fig:starttimes} compares, for each considered technology, two distributions of the introduction time of the technology in a project. The first distribution (left) considers the first time a technology gets introduced in a project. The second distribution (right) considers the introduction of the technology in a project that already had a technology before.
As expected, we observe that \textbf{more than 50\% of the introductions of a first technology are done in the first 10\% of the project's lifetime}. For technologies introduced after an existing one, the distribution tends to be flatter.

We also observe that the two distributions for \jdbc present less differences than the ones related to \jpa or \hbm. 
To achieve this, we performed a Kolmogorov-Smirnov statistical test for each pair of distributions related to \jdbc, \jpa and \hbm. 
The tests show that \textbf{the two distributions associated to each technology are significantly different} (p-values are lower than $10^{-6}$).
This may indicate that \textbf{for \jdbc, the moment of introduction is less affected by the presence of another technology than for \hbm and \jpa}. 

We saw that the time at which a technology is introduced in a project varies depending on the presence of another technology in this project. 
What are the technologies that are more likely to be succeeded by another one?

To answer this question, we use the statistical technique of \emph{survival analysis} to estimate the probability that a technology does not remain the last introduced one in a project lifetime. 
Survival analysis \cite{Samoladas2010} creates a model estimating the survival rate of a population over time, considering the fact that some elements of the population may leave the study, and for some other elements the event of interest does not occur during the observation period.
In our case, the observed event is the introduction in a project of another technology after an existing one.  

\begin{figure}[!htbp]
\centering
\includegraphics[width=0.9\columnwidth]{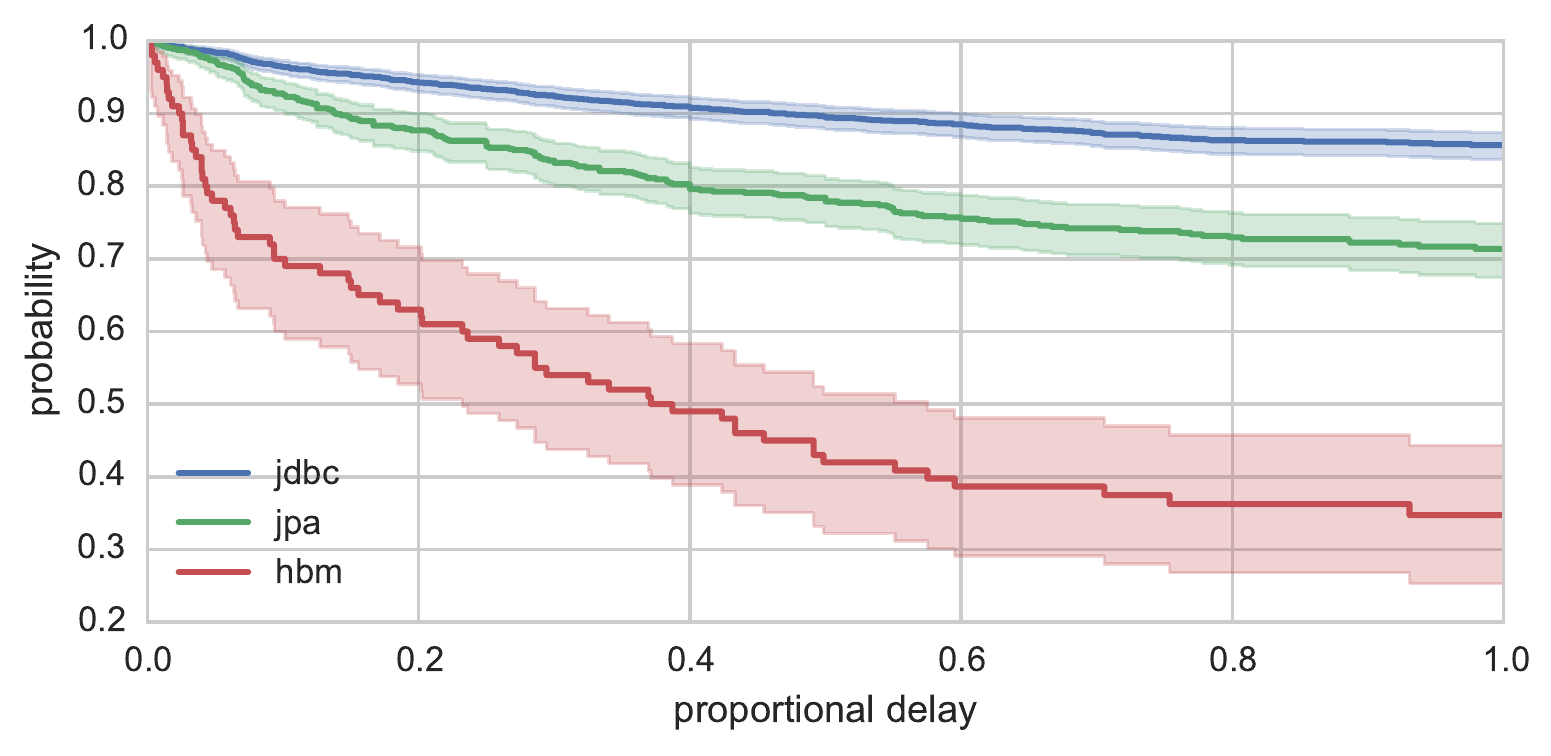}
\caption{Probability that a technology remains the last introduced technology over time.\label{fig:last_introduced}}
\end{figure} 

Figure~\ref{fig:last_introduced} shows the survival rates for each considered technology. 
We observe that \hbm has a much lower survival rate (\ie a lower probability of staying the last introduced technology for a long time) than the other technologies.
We also observe that, during the first 10\% of the projects' lifetime, the survival rates
of \hbm decrease by 30\%, representing a more important decrease than for the other two technologies. This implies that \textbf{\hbm is usually quickly replaced or complemented by another technology}.

Figure~\ref{fig:nb_by_techs} showed that around 23\% of the projects use two or more database technologies in their lifetime, but these are not necessarily used simultaneously. 
We therefore identified which combinations of technologies actually \emph{co-occur} in the selected Java projects. Frequent co-occurrences would reveal which technologies are complementary, and which technologies are used as supporting technologies of other ones.
For each pair of technologies, we counted the number of projects in which these technologies actually co-occur,  and in which order they were introduced in these projects.
The results are summarised in Table~\ref{tab:nb_by_techs}.

\begin{table}[!htbp]
\centering
{\footnotesize
\begin{tabular}{r|r|r|r}
     $(A,B)\rightarrow$ & (\jdbc, \jpa) & (\jdbc, \hbm) & (\jpa, \hbm) \\
     \hline
       \# projects   & 497& 152 & 84\\
      \# co-occurrences  & 488 & 148 & 77  \\
      \% co-occurrences & 98.2\% & 97.4\% & 91.7\%  \\
      $\textsf{start}_A<\textsf{start}_B$  & 157& 50&19 \\
     $\textsf{start}_A>\textsf{start}_B$ & 151&27 &37 \\
     $\textsf{start}_A = \textsf{start}_B$ & 189& 75& 28\\
\end{tabular}
}
\caption{Projects characteristics by pairs $(A,B)$ of co-occurring technologies\label{tab:nb_by_techs}}
\end{table}

\textbf{Among all projects that use multiple technologies during their lifetime we observe a very high proportion of co-occurring technologies}. 
More specifically, in 97.3\% (488+148+77 out of 497+152+84) of all the situations in which two distinct technologies were used during a project's lifetime, they were used simultaneously. 
Around 41\% (189+75+28 out of 488+148+77) of all pairs of co-occurring technologies were introduced simultaneously ($start_A = start_B$), implying that around 59\% of all pairs of co-occurring technologies concern projects in which the technologies were introduced at different moments ($start_A \neq start_B$). 

Considering the number of projects in which the introduction of a technology $A$ was observed before the use of a technology $B$, it seems that \textbf{\jpa tends to succeed to \hbm more often than the contrary} (37 versus 19 observations).
Similarly, \textbf{\hbm tends to succeed to \jdbc more often than the contrary} (50 versus 27 observations).
We did not identify such an order for \jpa and \jdbc (151 versus 157 observations).

\begin{mdframed}
\textbf{Summary.}
All considered technologies are introduced early in the projects' lifetimes, even for projects that already use another technology. 
The number of projects in which multiple technologies co-occur is proportionally important.
The order in which these technologies are introduced suggests that \hbm is often succeeded by \jdbc or \jpa. 
\end{mdframed}


\section{$RQ_2$ How does the introduction of a new technology in a project affect the already included ones?}

As multiple database access technologies are used in many projects, either simultaneously or one after the other, it is useful to study how the introduction of a new technology can impact the use of an already included one. 
This impact, if it occurs, could result in an increased or decreased usage of the already included technology.
We therefore identified and counted for which projects the introduction of a new technology causes an increasing use of the older technology, a decreasing use, or no observable change in the use of the already included technology.

To qualify the impact, we rely on the \emph{first derivative} of the number of files related to an existing technology.
We computed and compared the mean of this derivative for two 8-week periods: the first period strictly precedes the moment of introduction of the new technology, and the second period immediately follows the moment of introduction. 

In the following, we will use the term {\em variation} to denote the difference between the mean of the second period and the mean of the first period.  
The variation of a technology is easy to interpret: a positive value indicates an increasing use of the existing technology while a negative value indicates a decreasing use of the existing technology

\begin{figure}[!htbp]
\centering
\includegraphics[width=\columnwidth]{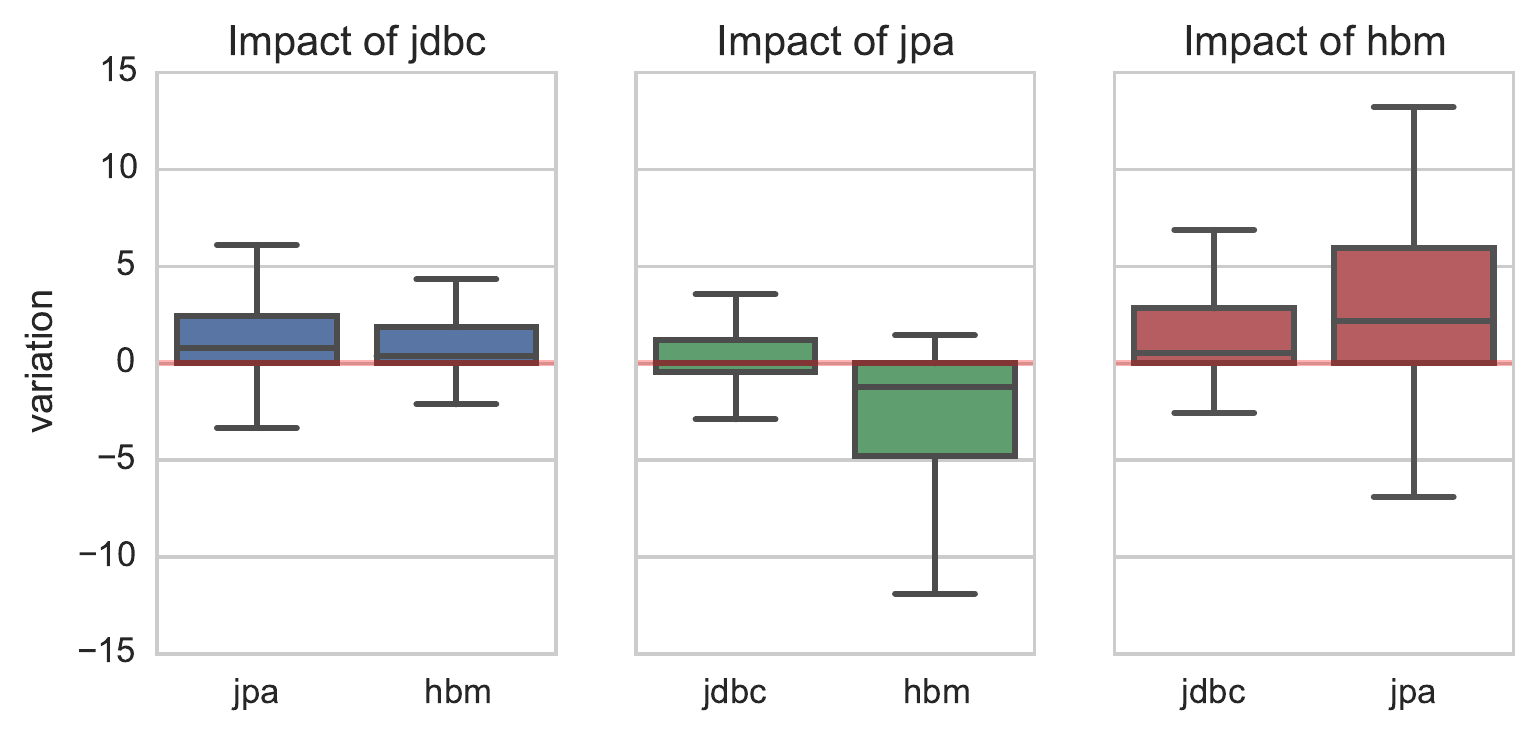}
\caption{Impact of the introduction of a new technology on the activity of an already included technology.\label{fig:impact_value}}
\end{figure}

Figure~\ref{fig:impact_value} shows the distribution of the variation for each pair of technologies. 
We observe that \textbf{\jdbc and \hbm cause a slight positive impact on the use of existing technologies} (since the variation tends to be positive in 75\% of all cases). 
Notice the important variation induced by introducing \hbm in projects using \jpa.
The converse is not true: \textbf{introducing \jpa in a project that already uses \hbm implies a negative variation for \hbm}. 

\begin{figure}[!htbp]
\centering
\includegraphics[width=\columnwidth]{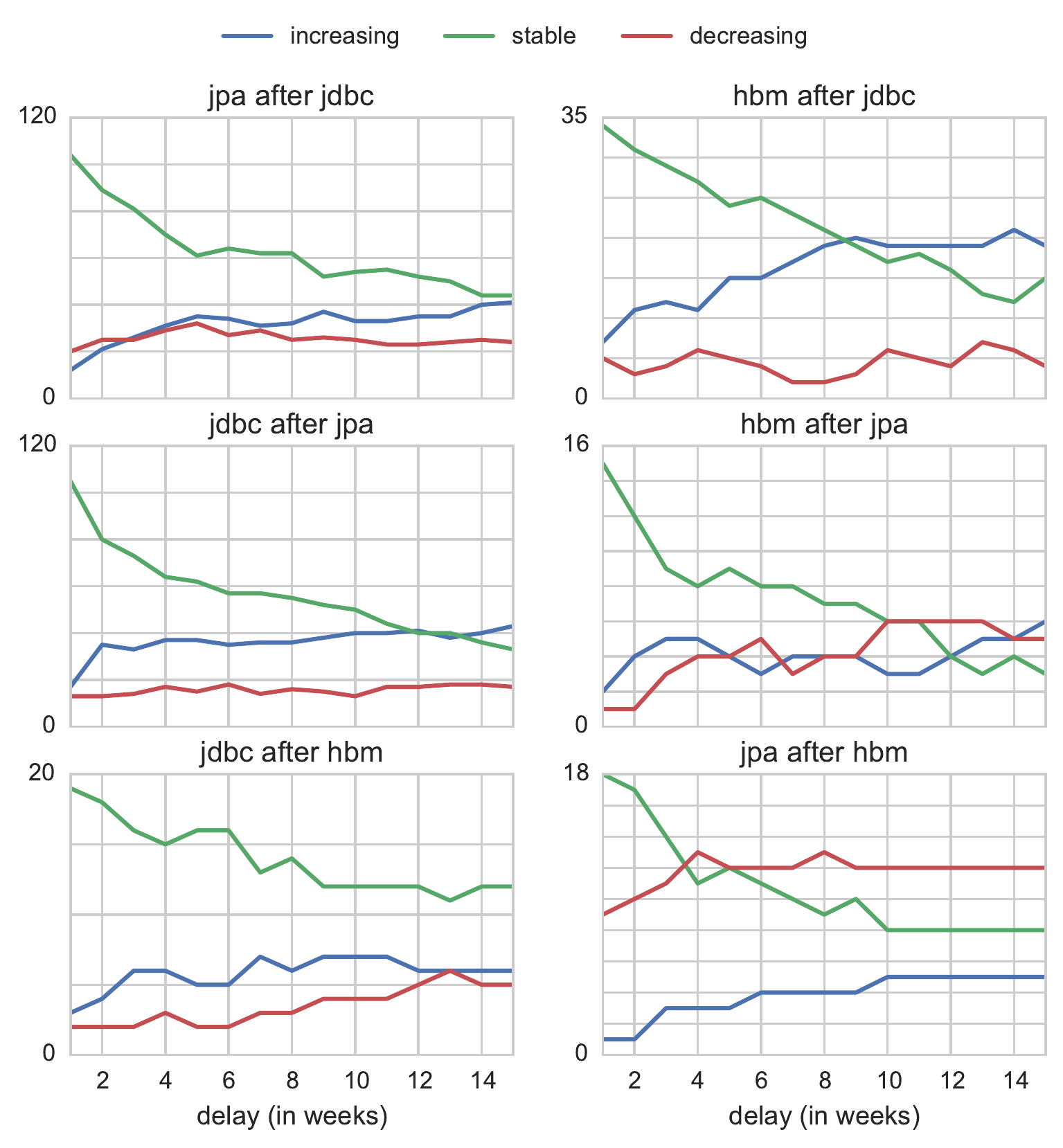}
\caption{Number of projects with an increasing, decreasing or stable activity of an already included technology, as observed \textsf{x} weeks after introducing another technology.\label{fig:impact_number}}
\end{figure}

Figure~\ref{fig:impact_value} only identifies global trends in our project corpus. It does not allow to identify trends within individual projects. 
Figure~\ref{fig:impact_number} therefore distinguishes the projects that exhibit a positive variation (\textbf{\color{darkblue}blue curve}), a negative variation (\textbf{\color{darkred}red curve}) or no variation (\textbf{\color{darkgreen}green curve}) for several time intervals after the introduction of the new technology. 

Regardless of the considered pair of technologies, with the notable exception of the pairs (\jpa after \hbm) and (\hbm after \jpa), both the number of projects having no variation and the number of projects having a positive variation are systematically greater than the number of projects exhibiting a negative variation. 

\begin{figure}[!htbp]
\centering
\includegraphics[width=\columnwidth]{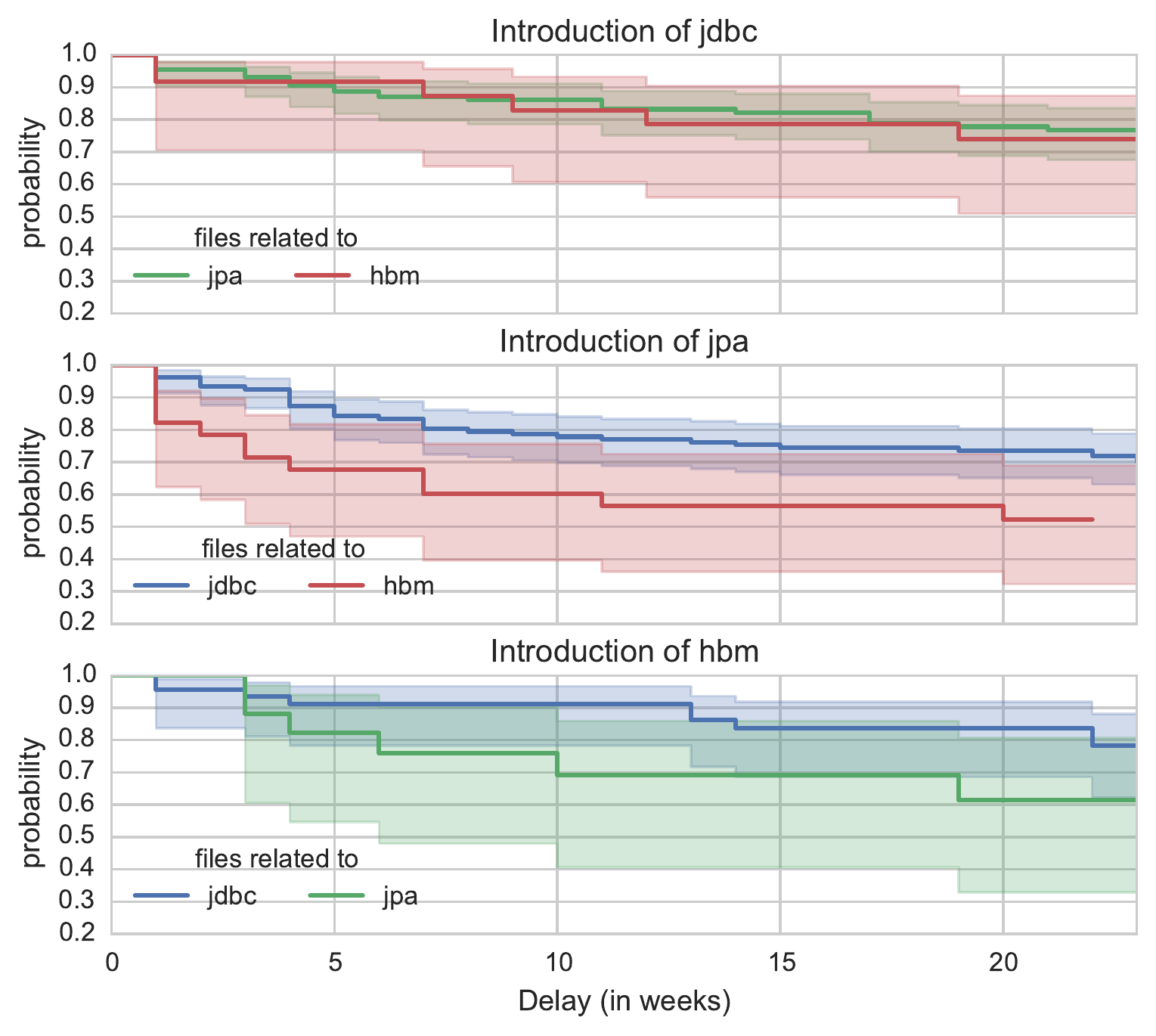}
\caption{Probability that at least 25\% of files related to a technology remain after the introduction of another technology.\label{fig:impact_remaining}}
\end{figure}

Figure~\ref{fig:impact_remaining} shows survival curves, using a Kaplan-Meier estimator, of the probability that a project keeps more than a threshold of $25\%$ of its files related to an already included technology after the introduction of new one.
We tried different threshold values and they all lead to the same conclusions.

Again, we observe that \textbf{the most distinct behaviours are exhibited by \jpa and \hbm}: the probability to keep more than $25\%$ of files related to \hbm drops below $0.55$ about 20 weeks after introducing \jpa, while the probability for \jpa files drops to a little more than $0.6$ about 19 weeks after introducing \hbm.
This analysis corroborates our previous observations: \textbf{introducing \jpa or \hbm does not negatively impact the use of \jdbc}, and conversely.
We also observe from Figures~\ref{fig:impact_number} and~\ref{fig:impact_remaining} that \textbf{most of the impact happens in the first weeks after introducing the new technology}. 

\bigskip

\begin{mdframed}
\textbf{Summary.}
Introducing a new technology generally induces, in the short term, an increase of the presence of the already included technology, with the notable exception of the introduction of \jpa on a project that already makes use of \hbm.  
This suggests that, contrary to the promises of ORM technologies, new technologies do not tend to replace existing ones but rather complement them.
\end{mdframed}



\section{$RQ_3$ To which extent does the introduction of a technology impact the way in which a project accesses the database?}

From the results of $RQ_1$ we observed that, if a project uses multiple database access technologies over its lifetime, these technologies tend to co-occur. At a more fine-grained level, we are interested in the impact of the introduction of a technology on the files that already relate to a previously used technology. 

\subsection{Do different technologies co-occur at file level?}

\begin{figure}[!htbp]
    \centering
     \includegraphics[width=\columnwidth]{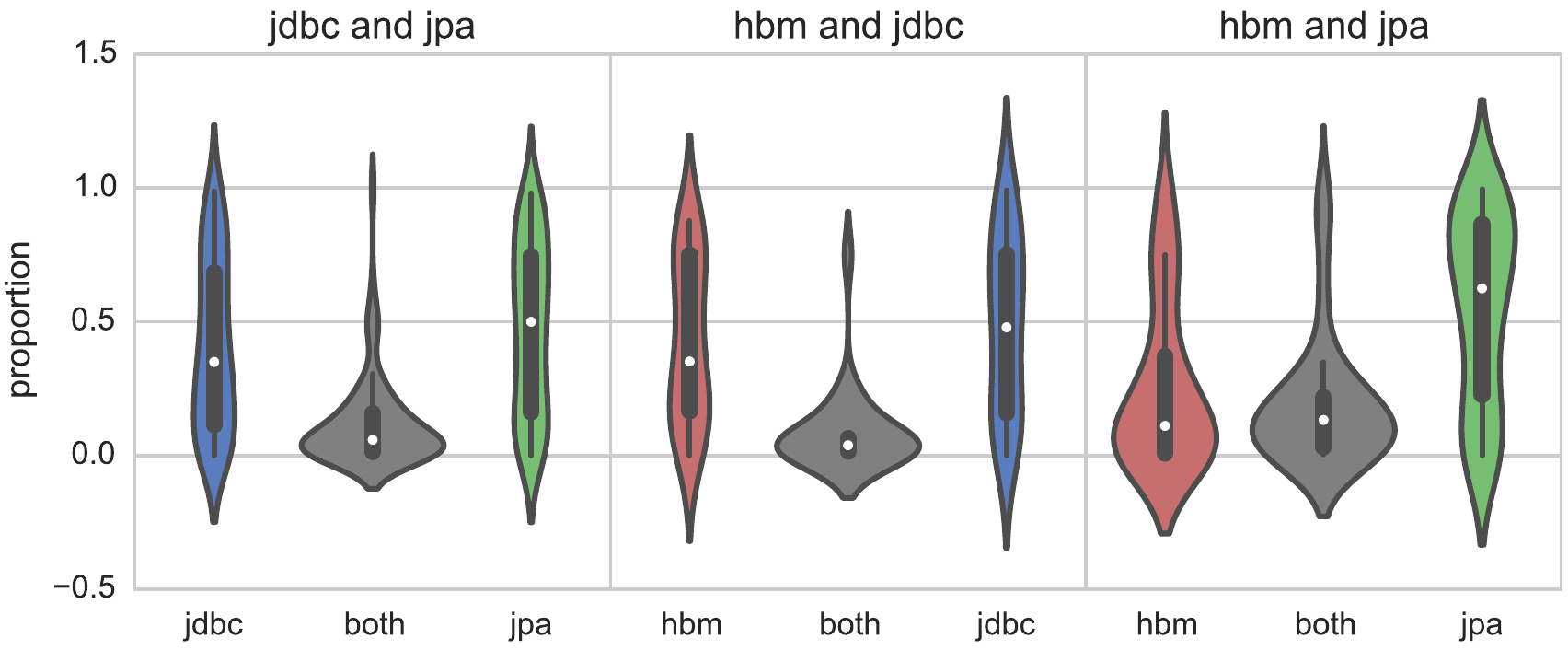}
     \caption{Relative number of files relating to pairs of technologies.\label{techno-relative-purity}}
\end{figure}

Let us  first study the co-occurrences of different technologies at file level without taking the evolutionary aspect into account.
Figure~\ref{techno-relative-purity} shows, for each pair of technologies, the distribution across projects of the ratio between the number of files that relate to each, or both, technologies, and the number of files that relate to any of these technologies. For each pair of technologies, only projects in which both technologies have been used at some point in their lifetime have been retained as elements of the distribution. 

It turns out that pairs of technologies including \jdbc present similar profiles: most projects contain a small proportion of files using both technologies. A two-sided Kolmogorov-Smirnov test confirms this similarity between distributions: we cannot reject the null hypothesis that states that the distributions associated to the proportion of files using a single technology are identical ($p = 0.877$ and $0.287$, respectively). We conclude that \textbf{\jdbc is generally not used in the same files as \jpa and \hbm}.

The pair of technologies \jpa and \hbm presents a different behaviour. 
The three distributions of the proportion of files that only relate to these technologies are significantly different (we reject the null hypothesis with $p < 0.001$).
This result, combined with the form of the distributions, suggests that, for projects having used \jpa and \hbm, \textbf{a file is likely to relate either to \jpa only or to both \jpa and \hbm}. 
In addition to this, the proportion of files that use both \hbm and \jpa is more important than for the other considered pairs of technologies.

\bigskip

\begin{mdframed}
\textbf{Summary.}
There is a clear separation between files using \jdbc and files using the two other technologies. For the combination of \hbm and \jpa, a partial, asymmetric overlap exists at file level: \hbm is often used in the same files as \jpa, while \jpa is rarely used in combination with another technology in the same file.
\end{mdframed}

\subsection{How does the co-occurrence of technologies at file level evolve over time?}

Let us now look at the same question from an evolutionary point of view, by assessing
the impact, at file-level, of introducing a new technology in a project that already uses another technology to access the database. To do this, we study how the files related to an existing technology get changed after introduction of the new technology. 

Let us associate a \emph{migration profile} to each project at different points in time after the introduction of the new technology. This migration profile reflects how the files related to the old technology are impacted. It is computed as follows:

Let $P$ be a project and $\mathcal{T} = \{ \jdbc,\hbm, \jpa \}$ the considered technologies.
For each point in time $t$ for $P$ and each technology $A \in \mathcal{T}$ we define $\related{A}{t}$ as the (possibly empty) subset of (fully qualified) filenames of $P$ in which technology $A$ was detected at time $t$. 

For every pair of distinct technologies $(A,B) \in \mathcal{T}\times\mathcal{T}$, we write $M = (P,A,B)$ if $P$ is a project in which technology $B$ gets introduced while a technology $A$ is already in use. Let $t_M$ denote the point in time of this introduction and $F_M = \related{A}{t_M}$ the set of filenames associated to technology $A$.
For each $ t\geq t_M$ we associate to each $f \in F_M$ a label in $\mathcal L = \{ \resid, \suppr, \compl, \repl \}$ as follows:

   \resid if $f\in\related{A}{t} \setminus \related{B}{t}$
   
   \suppr if $f\notin\related{A}{t} \cup \related{B}{t}$
   
   \compl if $f\in\related{A}{t} \cap \related{B}{t}$
  
   \repl if $f\in\related{B}{t} \setminus \related{A}{t}$

Given $M$, we also associate to each $t\geq t_M$ a set of labels $\profiles{t} \subseteq \mathcal L$. 
A label $L\in\mathcal L$ belongs to $\profiles{t}$ if, among the labels associated to each $f \in F_M$ at time $t$, no other label occurs more frequently than $L$.

Finally, the \emph{migration profile} of $M$ at time $t$ is a unique label from $\profiles{t}$ selected  based on the total order $\repl > \compl > \suppr > \resid$.
This total order privileges migration profiles that correspond to the adoption of the new technology.

As the choice of a total order could have altered the results of our analysis, we compared the results obtained with several total orders, and we observed only slight local variations.
This is not surprising as there are only 72 pairs $(\mathsf M, t)$ such that $|\profiles{t}| > 1$, representing 1.78\% of all the considered pairs. 

\begin{figure}[!htbp]
    \centering
    \includegraphics[width=\columnwidth]{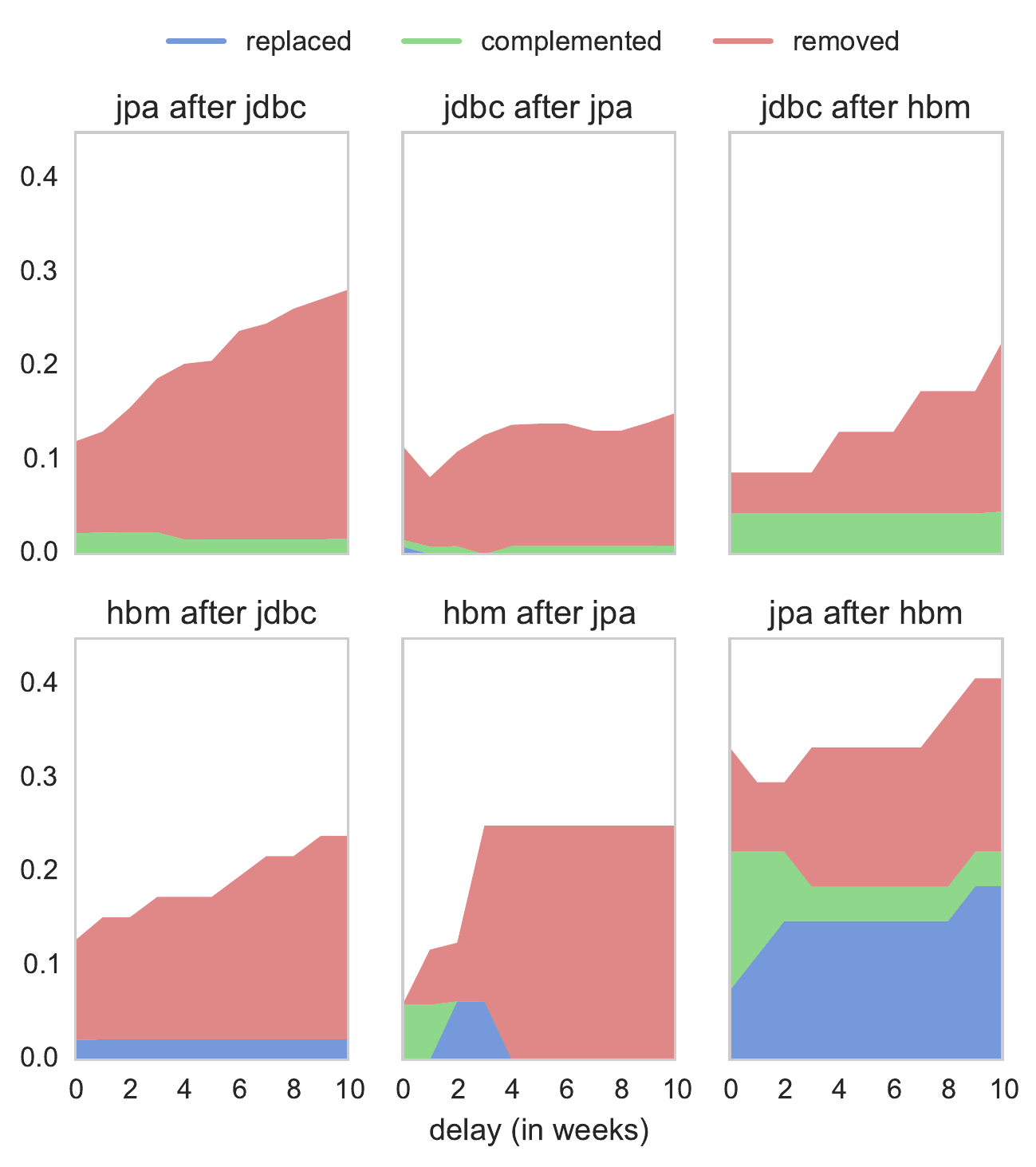}
    \caption{Proportion (stacked) of projects for each migration profile. The complement corresponds to \repl. \label{fig:existing-file-contamination}}
\end{figure}

Figure~\ref{fig:existing-file-contamination} shows the evolution of the proportion of projects with a given migration profile. 
For the sake of readability, we only present results for \compl, \repl, and \suppr. 
The results for \resid can be deduced from these, by taking the complement of \compl, \repl and \suppr. 

We observe that, for each considered pair of technologies, and for each time delay (expressed in weeks) after the introduction of the new technology, most projects relate to the \resid migration profile, implying that projects tend not to adapt their existing database access files to make use of the newly introduced technology. 
This is especially true for projects introducing \jdbc after \jpa or \hbm. 

The second dominant migration profile is \suppr. 
Regardless of the considered pair of technologies, more and more projects are associated to this migration profile. 
Over time, an increasing number of projects tend to reduce the number of files relating to the first considered technology. 
The predominance of \resid and \suppr migration profiles seems to convey that, \textbf{in many cases, files that related to the existing technology are not prone to use the newly introduced technology.}
Instead, they either continue to use the first technology or they tend to lose any relation to database access management.

The two other migration profiles, \compl and \repl, indicate an effective file migration from the existing technology to the newly introduced one.
Such cases appear to be much less represented in our corpus, with the exception of projects in which \jpa or \jdbc is introduced after \hbm. This is especially the case \textbf{when \jpa is introduced in a project using \hbm: the files that were related to \hbm become (sometimes exclusively) related to \jpa. }

\smallskip
\begin{mdframed}
\textbf{Summary.}
Different technologies generally do not tend to co-occur in the same set of files, except, to some extent, when \jpa and \hbm are used together.
We do not observe a true migration in technology usage:  files that are related to a given technology do not tend to adopt the newly introduced technology, except for projects that migrate from \hbm to another technology.
\end{mdframed}


\section{Threats to validity}
\label{sec:threats}

Our research suffers from the same threats as other research relying on Git and GitHub~\cite{bird_promises_2009, DBLP:conf/msr/KalliamvakouGBSGD14}. 

The selected Java projects potentially suffer from the same generalisability constraints as in \cite{githubCorpus2013}. The open source GitHub Java project corpus was curated to exclude low-quality projects (by ignoring projects that were never forked) and project duplicates.

While our corpus contained 2,457 projects, the number of projects involved in some pairs of database technologies were sometimes much lower. For example, only 19 projects were concerned by a migration from \jpa to \hbm (cf. Table~\ref{tab:nb_by_techs}).
The accuracy of our observations could be increased by using a larger project corpus. 

The detection of a technology is based on the static analysis of code and project-specific artefacts (\eg Java annotations, import statements and XML files).
This approach can lead to false positives: the presence of these artefacts does not necessarily reflect the actual use of the related technology.

Some of our analyses are based on arbitrarily chosen thresholds and on weekly time intervals.
Because our results may depend on these thresholds and intervals, we repeated our experiments with different parameters but did not observe any major differences. 


\section{Future Work\label{sec:future-work}}

The results presented in this article, possibly combined with more traditional project quality metrics, could be integrated in a managerial dashboard.
Such a dashboard could be used to compare the characteristics and the evolution of a particular project against those belonging to the analysed project corpus. This would support project managers in evaluating and exploiting the expected benefits and disadvantages from introducing a new technology, as well as in assessing the impact of how this technology will become used in the project over time. Any ensuing managerial decisions will obviously depend on project-specific rules and  guidelines that could hardly be generalized.

This paper used static analysis techniques to detect the presence of a particular technology. Using dynamic analysis techniques could reveal how database technologies are actually used in running systems. The analysis of queries submitted to the database at runtime could be used for understanding to which extent ORM technologies hide complexity to developers.   

This paper focused on relational database access technologies based on three representative technologies (\jdbc, Hibernate and \jpa). It could be useful to include other Java specifications for object persistence as well, such as JDO. It would also be useful to consider other kinds of databases (such as NoSQL, graph or object-oriented databases), since these are becoming increasingly more popular. A follow-up study could take into account such alternative database technologies.

Other technological domains (beyond databases) could be considered as well. Event loggers, graphical user interfaces, and unit tests are examples of features supported by multiple concurrent technologies. Since the identification of the technology used in project files is the only part of our methodology that depends on the considered technologies, our approach could be easily adapted to study other technologies.

\sect{sec:threats} mentioned the limitations of the selected project corpus. We therefore intend to confirm our research results by considering a larger project corpus, including both open and closed source projects. We also intend to study the effect of project quality and project maturity on the obtained results. Finally, we intend to include other programming languages than Java in the project corpus in order to avoid any bias introduced by language-specific characteristics.

While this paper only focused on \emph{technical} aspects of connecting source code to databases, we plan to study the \emph{social} aspects of systems involving such a database connection. More precisely, we would like to determine if the different technologies are introduced and managed by different teams or persons. Inspired by~\cite{Vasilescu2013} we also aim to analyse the developer characteristics in order to determine how these affect the take-up, use, evolution and migration of technologies. Some examples of developer characteristics are their degree of specialisation, diversity, seniority, skills, and workload.


Finally, we plan to analyse software systems in order to automatically identify library features used in the source code, as well as feature similarities between different technologies. In situations where developers want to migrate from a given technology to another, such a feature identification and mapping is a first step towards better support for assisted or automatic migration~\cite{teyton-function-mappings-discovery}.


\section{Conclusions\label{sec:conclusion}}

Through static analysis of Java source code we carried out a large-scale empirical study to understand how database access technologies interact with one another.
We considered three popular technologies (\jdbc, \hbm and \jpa) that represent different means to connect Java source code files to a relational database. We selected data from 2,457 open source projects on GitHub that used at least one of the considered technologies.

Our study revealed common behaviours in the use of these three technologies. In spite of the promises of ORM technologies, we found no evidence that the low-level \jdbc solution is massively replaced by \hbm or \jpa. The only significant technology migration we observed concerns the transition from \hbm to \jpa.   
More specifically, we summarise our main observations below.

We analysed the evolution and co-occurrences of the technologies in order to get a high-level view of their usage in the considered Java projects. It appears that, most of the time, database technologies are introduced early in the projects' lifetime, whether they are the first technology introduced or not. Once introduced in a project, \hbm tends to be complemented or replaced by another technology more frequently and more quickly than \jpa and \jdbc.

We also analysed how the technologies are used in the source code files. The introduction of \jdbc and \hbm tends to be followed by an increasing use of the already present database technology. This increase is particularly important when \hbm is introduced after \jpa. Conversely, the introduction of \jpa reduces the use of \hbm. \jpa therefore appears to replace existing \hbm in the database-related source-code files, while the converse is not true. 

Furthermore, \jdbc generally does not share source code files with the two other considered database technologies. While \jpa is used in isolation in a majority of source code files, \hbm tends to be used more often in conjunction with \jpa. The study of the evolution of such co-occurrence reveals that a file migration from a technology to another one is only observed from \hbm to \jpa. In most projects, the introduction of a new database technology is not followed by a massive adoption of this technology by the existing database-related files, until these files become database-unrelated or are removed from the source code repository.

Exploiting all these results in a dashboard that supports managers in making project-specific decisions with respect to the introduction, use or evolution of database access technologies remains part of future work. 


\section*{Acknowledgment}
This research was conduced as part of the FRFC research project T.0022.13 ``Data-Intensive Software System Evolution'' that was financed by the F.R.S.-FNRS, Belgium.


\newpage
\providecommand{\noopsort}[1]{}

\end{document}